\documentclass{svproc}
\usepackage{tikz}
\usetikzlibrary{arrows,shapes}
\usetikzlibrary{arrows.meta}

\usetikzlibrary{shapes.callouts,decorations.pathmorphing}
\usetikzlibrary{calc,positioning}

\def\ubr#1#2{\underbrace{#1}_{\text{#2}}}

\usepackage{amssymb}
\usepackage{amsmath}
\usepackage{url}

\tikzstyle{myrectangle}=[rectangle,rounded corners, fill=white, 
draw=uibred,align=center,line width=1pt]
\tikzstyle{myrectangleg}=[rectangle,rounded corners, fill=white, 
draw=MyDarkGreen,align=center,line width=1pt]
\tikzstyle{myarrow}=[->,>=stealth,draw=uibred,line width=3pt]
\tikzstyle{mycircle}=[fill=white, fill opacity = 0.8, draw, line width=1pt]
\tikzstyle{myfigure}=[anchor=south west,fill=white]

\begin{document}
\mainmatter              

\title{Chemical equilibration of QGP in hadronic collisions}

\author{Aleksi Kurkela\inst{1,2} \and Aleksas Mazeliauskas\inst{1,3}}
\institute{
Theoretical Physics Department, CERN, 1211 Gen\`eve 23, Switzerland
\and
Faculty of Science and Technology, University of Stavanger, 
4036 Stavanger, Norway
\and
Institut f\"{u}r Theoretische Physik, Universit\"{a}t Heidelberg, 
69120 Heidelberg, Germany
}

\maketitle              

\begin{abstract}
We performed state-of-the-art QCD effective kinetic theory simulations of chemically equilibrating QGP in longitudinally expanding systems. We find that chemical equilibration takes place after hydrodynamization, but well before local thermalization. By relating the transport properties of QGP and the system size we estimate that hadronic collisions with final state multiplicities $dN_\text{ch}/d\eta > 10^2$ live long enough to reach approximate chemical equilibrium for all collision systems. Therefore we expect the saturation of strangeness enhancement to occur at the same multiplicity in proton-proton, proton-nucleus and nucleus-nucleus collisions.
\end{abstract}
The experimental measurements of the final-state particles in the ultra-relativistic proton-proton, proton-nucleus and nucleus-nucleus collisions at 
hadron colliders show a qualitative change of hadron chemistry with the increasing particle multiplicity~\cite{ALICE:2017jyt,Abelev:2013haa,Abelev:2013zaa}. At central nucleus-nucleus collisions the observed 
ratios of long-lived hadrons is consistent with chemical equilibrium hadron
resonance gas models at temperature $T_\text{ch}\approx
155\,\text{MeV}$~\cite{Andronic:2017pug}, while the lowest multiplicity
proton-proton collisions are reproduced by perturbative event generators. One
interpretation behind this change is the formation of deconfined state of
Quark-Gluon Plasma (QGP), which reaches approximate thermal and chemical equilibrium
in the collision fireball, from which the thermal hadron productions ensues. 
Indeed, the hydrodynamic models of the QGP expansion have been very successful in describing small transverse momentum particle spectra and multi-particle correlations.
However the initial state created in high-energy nuclear collisions even locally is far from equilibrium and
the study of thermalization has been a very active topic~\cite{Schlichting:2019abc}.
In this work~\cite{Kurkela:2018xxd,Kurkela:2018oqw}
 we describe the QGP approach to chemical equilibrium within the framework of QCD effective
kinetic theory~\cite{Arnold:2002zm}.

At high energy, weak coupling limit of QCD, the mid-rapidity interaction region is
dominated by strong non-equilibrium gluonic fields~\cite{Schlichting:2019abc}. The microscopic
description to local thermal
equilibrium from such initial configuration 
is described by the ``bottom-up" thermalization scenario~\cite{Baier:2000sb}, which
was explicitly realized by classical-statistical Yang-Mills~\cite{Berges:2013eia} and gluonic kinetic theory~\cite{Kurkela:2015qoa} simulations. Here for the first time we use the complete leading order QCD kinetic
theory of quarks and gluons~\cite{Arnold:2002zm} to simulate the hydrodynamic, chemical and kinetic equilibration of QGP.
We solve the coupled Boltzmann equations for quark and gluon distribution functions undergoing
homogeneous boost-invariant expansion
\begin{align}
\partial_\tau f_{g,q} - \frac{p_z}{\tau}\partial_{p_z} f_{g,q}= 
-\mathcal{C}_{2\leftrightarrow2}[f]
- 
\mathcal{C}_{1\leftrightarrow2}[f].
\end{align}
We include elastic $2\leftrightarrow2$ and colinear $1\leftrightarrow2$ collision processes
at leading order in the coupling constant $\lambda=4\pi \alpha_s N_c$. We consider $N_c=3$ colors
and  $N_f=3$ flavours of massless fermions with equal quark-antiquark content, i.e., at vanishing
baryon chemical potential $\mu_B=0$.
For the explicit expressions of collision kernels and the regularization procedure of the infrared divergences see the published work~\cite{Kurkela:2018oqw}.

Following the previous work~\cite{Kurkela:2015qoa}, initially at $\tau = 1\, Q^{-1}_s$ the fermion distribution function $f_q=0$ is set to zero, while 
the initial gluon density is
\begin{align}
{f}_{0}=\frac{2 A}{\lambda} \frac{Q_0}{\sqrt{p_\perp^2+p_z^2\xi^2}} e^{-\frac{2}{3} \frac{p_\perp^2+\xi^2 p_z^2}{Q_0^2}},\label{eq:focc}
\end{align}
where the values of $A=5.24$ and $Q_0=1.8\,Q_s$ are adjusted to match transverse momentum and energy density extracted from the lattice. The parameter $\xi$ determines the initial anisotropy and is set to $\xi=10$ to reflect highly anisotropic initial
conditions in heavy ion collisions. We then follow the evolution of quark and gluon distribution functions to
equilibrium at different values of the coupling constant $\lambda$ and find the corresponding values of 
shear viscosity over entropy ratio $\eta /s$. In the following we discuss the time evolution of energy density
obtained by momentum integration of the non-equilibrium quark and gluon distribution functions.

It was observed in previous works~\cite{Kurkela:2015qoa,Keegan:2015avk,Kurkela:2018vqr,Kurkela:2018wud}, that 
the parametrically large differences in the equilibration rates for different coupling constants $\lambda$ can be largely scaled out
by measuring time in units of \emph{relaxation time}
\begin{equation}
  \tau_R(\tau) = \frac{4\pi \eta/s}{T_{\rm id.}(\tau)},\label{eq:tauR}
\end{equation}
where $\eta/s$ is the specific shear viscosity. Note that for expanding system the
local thermalization temperature and the relaxation time itself is time dependent. Here we define the effective temperature $T_{\rm id.}(\tau) = \frac{(T(\tau)\tau^{1/3})|_{\tau \rightarrow \infty}}{\tau^{1/3}}$,
which asymptotically coincides with the temperature of near equilibrium QGP.

\begin{figure}[t]
\centering
\includegraphics[width=0.5\linewidth]{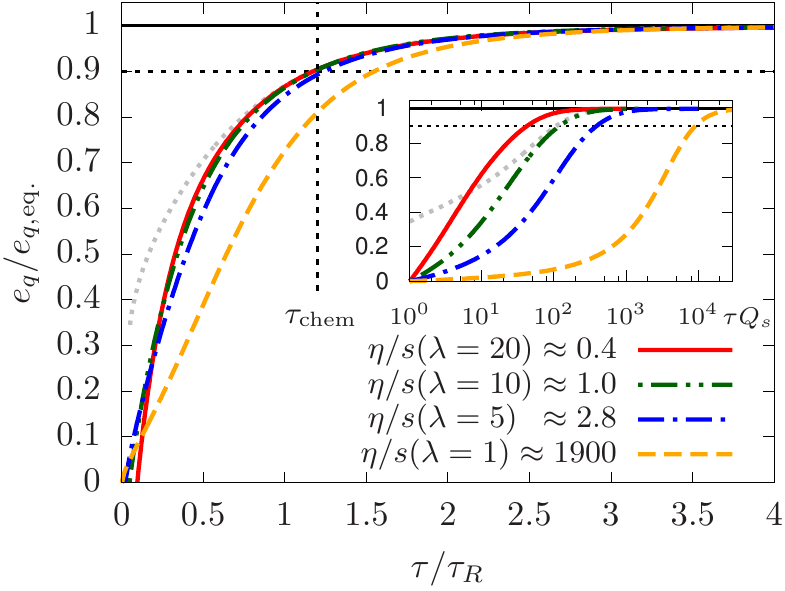}%
\includegraphics[width=0.5\linewidth]{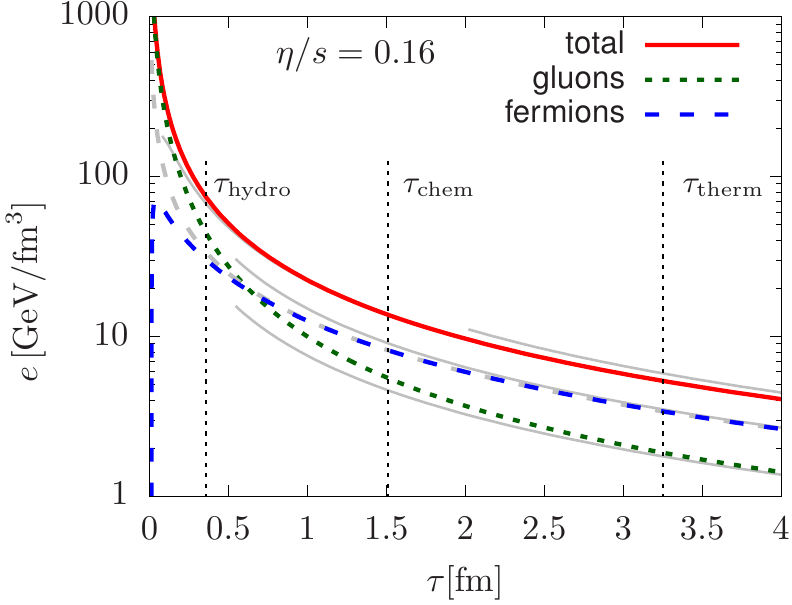}
\caption{\label{fig1} a) 
Fermion energy density  fraction $e_q(\tau)/e_{q,\text{eq}}(\tau)$ as a function of rescaled time for different coupling constants.  The inset shows un-rescaled time dependence.
 b)
Evolution of total energy density and its gluonic and fermion components in kinetic theory
 converted to physical units using universality of $\tau/\tau_R$ scaling and 
phenomenological values of $\eta/s=0.16$, $\left<s\tau\right>=4.1\,\text{GeV}^2$ and $\nu_\text{eff}=40$. The grey lines 
correspond to ideal, viscous and chemically equilibrated energies.
 Figures taken from Ref.~\cite{Kurkela:2018xxd}
}
\end{figure}

In chemical equilibrium, $N_f=3$ fermions constitute $e_{q,\text{eq}}/e_{\text{total}}\approx 0.66$  of the total equilibrium energy density. In Fig.~\ref{fig1}(a) we present the time evolution of quark energy density relative to equilibrium expectation as a function of rescaled time  time $\tau/\tau_R$. 
We consider different values of the coupling constant $\lambda=1,5,10,20$, which corresponds to $\eta/s\approx 1900, 2.8,1.0,0.4$ and vastly different relaxation times $\tau_R$. Remarkably, the large separation of equilibration
time-scales shown in the inset of Fig.~\ref{fig1}(a) collapses to near $\lambda$-independent equilibration
curve shown in the main panel of Fig.~\ref{fig1}(a). We find that  at the time $\tau/\tau_R \approx 1.2$ and for
larger values of the coupling, the 
fermions reach 90\% of their chemical equilibrium energy density. We checked that this is not affected my small non-zero fermion density at early times indicated by grey dashed line in  Fig.~\ref{fig1}(a).

We can convert the scaled chemical equilibration time $\tau_\text{chem} / \tau_R$ to physical time by solving Eq.~\eqref{eq:tauR} and
noting that the asymptotic constant $T^3_\text{id.}\tau = (T^{3}\tau)_\infty $ is proportional to
the local entropy density per rapidity $(s \tau)_\infty$:
\begin{equation}
  \tau_\text{chem} = \ubr{(\tau_\text{chem}/\tau_R)^{3/2}}{scaled time variable}\times \ubr{(4\pi\eta/s)^{3/2}\times ( s \tau)_\infty^{-1/2}\times (4 \pi^2 \nu_\text{eff}/90)^{1/2}}{phenomenological input}\label{eq:conv}.
\end{equation}
Substituting the empirical estimate $\tau_\text{chem}/\tau_R=1.2$ we obtain a pocket formula for
chemical equilibration time as a function of  entropy density, specific shear viscosity and number of degrees of freedom. 
It is important to emphasize that although
the parametric dependence of equilibration time on, say, entropy density, 
can be inferred from general physics arguments,
Eq.~\eqref{eq:conv} has all relevant numerical factors and is therefore a quantitative formula.
The approximate independence of $\eta/s$ seen in Fig.~\ref{fig1}(a), gives us motivation of plugging in phenomenological relevant parameter values of $\eta/s=2/4\pi\approx 0.16$, $\langle s\tau\rangle = 4.1 \, \text{GeV}^2$ and $\nu_\text{eff}=40$. In  Fig.~\ref{fig1}(b) we show the converted time evolution of total, gluon and fermion energy density. We observe that at $\tau\approx0.6\,\text{fm}$ 
fermion and gluon energy contributions become equal and at $\tau_\text{chem}\approx 1.5\,\text{fm}$ the chemical
equilibration defined above is achieved. In the paper~\cite{Kurkela:2018xxd,Kurkela:2018oqw}
we also studied hydrodynamic and kinetic equilibrations and found the following ordering of equilibration scales
(also shown in  Fig.~\ref{fig1}(b))
\begin{equation}
\ubr{\tau_\text{hydro}}{$\pm 10\%$ viscous $e(\tau)$} < \ubr{\tau_\text{chem}}{$\pm 10\%$ fermion eq. $e(\tau)$} < \ubr{\tau_\text{therm}}{$\pm 10\%$ ideal $e(\tau)$}.
\end{equation}
Strikingly, the empirical weak coupling equilibration formula Eq.~\eqref{eq:conv} produces realistic equilibration timescales compatible
with heavy-ion phenomenology if extrapolated to realistic values of $\eta/s$.
To what extent such extrapolation captures the dynamics of relatively strongly coupled system is debatable,
but it is clear that the weak coupling equilibration baseline is not in contradiction with rapid thermalization
of QGP.

Finally,  Eq.~\eqref{eq:conv} can be reformulated as a bound on charged particle multiplicities necessary to achieve chemical equilibrium
by freeze-out.
We relate
the multiplicity $dN_\text{\rm ch}/d\eta$  in terms of scaled variables using
$\langle\tau s\rangle \approx (S/N_\text{\rm ch})~1/(\pi R^2)~dN_\text{\rm ch}/d\eta$ (where 
$S/N_\text{\rm ch}\approx 7$~\cite{Hanus:2019fnc})
and  Eq.~\eqref{eq:conv}, so that
\begin{align}
\frac{dN_\text{ch}}{d\eta} =\tfrac{4\pi^3 }{90}\nu_\text{eff}\left({S}/{N_\text{ch}}\right)^{-1} \left(4\pi{\eta}/{s}\right)^3  \left({\tau}/{\tau_R}\right)^{3}  \left({\tau}/{R}\right)^{-2}.
\end{align}
Assuming that the system disintegrates once its lifetime exceeds the system size $\tau\sim R$~\cite{Kurkela:2018wud}, the minimum multiplicity needed for that time to be at or above chemical equilibration time $\tau_\text{chem}/\tau_R=1.2$ is 
\begin{equation}
 \frac{dN_{\rm ch}}{d\eta}  \gtrsim 110 \left(\frac{\tau_\text{chem}}{1.2\tau_R}\right)^3 \left(\frac{ \eta/s}{0.16}\right)^3\left(\frac{\tau_\text{chem}}{R}\right)^{-2}\label{eq:nch}.
\end{equation}
This bound is roughly compatible with experimentally observed trends of
strangeness enhancement~\cite{ALICE:2017jyt,Abelev:2013haa,Abelev:2013zaa},
which gives theoretical ground for the assumed formation of
chemically equilibrated QGP at high multiplicity collisions. 

In summary, we presented the detailed picture of QGP equilibration within QCD kinetic theory
framework. Strikingly, extrapolated to phenomenological values of 
$\eta /s$, this picture gives realistic equilibration timescales and
connects in a novel way the transport properties of QGP to experimental observations of hadron chemistry in
proton-proton, proton-nucleus and nucleus-nucleus collisions. 

\vspace{0.5cm}
\noindent
This work was supported in part by the German Research Foundation (DFG) 
Collaborative Research Centre (SFB) 1225 (ISOQUANT) (A.M.).

\bibliographystyle{./ProcSci_TeX/styles/bibtex/spmpsci_unsrt}
\bibliography{master}

\begin{thebibliography}{10}
\providecommand{\url}[1]{{#1}}
\providecommand{\urlprefix}{URL }
\expandafter\ifx\csname urlstyle\endcsname\relax
  \providecommand{\doi}[1]{DOI~\discretionary{}{}{}#1}\else
  \providecommand{\doi}{DOI~\discretionary{}{}{}\begingroup
  \urlstyle{rm}\Url}\fi

\bibitem{ALICE:2017jyt}
Adam, J., et~al.: {Enhanced production of multi-strange hadrons in
  high-multiplicity proton-proton collisions}.
\newblock Nature Phys. \textbf{13}, 535--539 (2017).
\newblock \doi{10.1038/nphys4111}

\bibitem{Abelev:2013haa}
Abelev, B.B., et~al.: {Multiplicity Dependence of Pion, Kaon, Proton and Lambda
  Production in p-Pb Collisions at $\sqrt{s_{NN}}$ = 5.02 TeV}.
\newblock Phys. Lett. \textbf{B728}, 25--38 (2014).
\newblock \doi{10.1016/j.physletb.2013.11.020}

\bibitem{Abelev:2013zaa}
Abelev, B.B., et~al.: {Multi-strange baryon production at mid-rapidity in Pb-Pb
  collisions at $\sqrt{s_{NN}}$ = 2.76 TeV}.
\newblock Phys. Lett. \textbf{B728}, 216--227 (2014).
\newblock \doi{10.1016/j.physletb.2014.05.052, 10.1016/j.physletb.2013.11.048}.
\newblock [Erratum: Phys. Lett.B734,409(2014)]

\bibitem{Andronic:2017pug}
Andronic, A., Braun-Munzinger, P., Redlich, K., Stachel, J.: {Decoding the
  phase structure of QCD via particle production at high energy}.
\newblock Nature \textbf{561}(7723), 321--330 (2018).
\newblock \doi{10.1038/s41586-018-0491-6}

\bibitem{Schlichting:2019abc}
Schlichting, S., Teaney, D.: {The First fm/c of Heavy-Ion Collisions}  (2019)

\bibitem{Kurkela:2018xxd}
Kurkela, A., Mazeliauskas, A.: Chemical equilibration in hadronic collisions.
\newblock Phys. Rev. Lett. \textbf{122}, 142,301 (2019).
\newblock \doi{10.1103/PhysRevLett.122.142301}

\bibitem{Kurkela:2018oqw}
Kurkela, A., Mazeliauskas, A.: {Chemical equilibration in weakly coupled QCD}.
\newblock Phys. Rev. \textbf{D99}(5), 054,018 (2019).
\newblock \doi{10.1103/PhysRevD.99.054018}

\bibitem{Arnold:2002zm}
Arnold, P.B., Moore, G.D., Yaffe, L.G.: {Effective kinetic theory for high
  temperature gauge theories}.
\newblock JHEP \textbf{01}, 030 (2003).
\newblock \doi{10.1088/1126-6708/2003/01/030}

\bibitem{Baier:2000sb}
Baier, R., Mueller, A.H., Schiff, D., Son, D.T.: {'Bottom up' thermalization in
  heavy ion collisions}.
\newblock Phys. Lett. \textbf{B502}, 51--58 (2001).
\newblock \doi{10.1016/S0370-2693(01)00191-5}

\bibitem{Berges:2013eia}
Berges, J., Boguslavski, K., Schlichting, S., Venugopalan, R.: {Turbulent
  thermalization process in heavy-ion collisions at ultrarelativistic
  energies}.
\newblock Phys. Rev. \textbf{D89}(7), 074,011 (2014).
\newblock \doi{10.1103/PhysRevD.89.074011}

\bibitem{Kurkela:2015qoa}
Kurkela, A., Zhu, Y.: {Isotropization and hydrodynamization in weakly coupled
  heavy-ion collisions}.
\newblock Phys. Rev. Lett. \textbf{115}(18), 182,301 (2015).
\newblock \doi{10.1103/PhysRevLett.115.182301}

\bibitem{Keegan:2015avk}
Keegan, L., Kurkela, A., Romatschke, P., van~der Schee, W., Zhu, Y.: {Weak and
  strong coupling equilibration in nonabelian gauge theories}.
\newblock JHEP \textbf{04}, 031 (2016).
\newblock \doi{10.1007/JHEP04(2016)031}

\bibitem{Kurkela:2018vqr}
Kurkela, A., Mazeliauskas, A., Paquet, J.F., Schlichting, S., Teaney, D.:
  {Effective kinetic description of event-by-event pre-equilibrium dynamics in
  high-energy heavy-ion collisions}.
\newblock Phys. Rev. \textbf{C99}(3), 034,910 (2019).
\newblock \doi{10.1103/PhysRevC.99.034910}

\bibitem{Kurkela:2018wud}
Kurkela, A., Mazeliauskas, A., Paquet, J.F., Schlichting, S., Teaney, D.:
  {Matching the Nonequilibrium Initial Stage of Heavy Ion Collisions to
  Hydrodynamics with QCD Kinetic Theory}.
\newblock Phys. Rev. Lett. \textbf{122}(12), 122,302 (2019).
\newblock \doi{10.1103/PhysRevLett.122.122302}

\bibitem{Hanus:2019fnc}
Hanus, P., Mazeliauskas, A., Reygers, K.: {Entropy production in pp and Pb-Pb
  collisions at the LHC}  (2019)

\end{thebibliography}

\end{document}